\documentclass[aps,pre,reprint,twocolumn,showpacs,showkeys,floatfix]{revtex4-2}
\usepackage{hyperref}
\usepackage{graphicx}
\usepackage{amsfonts}
\usepackage{amssymb}
\usepackage{epsfig}
\usepackage{epstopdf}
\usepackage{color}
\usepackage{bm}
\usepackage{pdfsync}

\newcommand{\be}{\begin{equation}}
\newcommand{\ee}{\end{equation}}
\newcommand{\bea}{\begin{eqnarray}}
\newcommand{\eea}{\end{eqnarray}}
\newcommand{\eul}{\mathrm{e}}
\newcommand{\kb}{k_{_{\mathrm{B}}}}

\newcommand{\qzero}{q_{_{\mathrm{0}}}}

\newcommand{\lambdaf}{\lambda_{_{\mathrm{f}}}}
\newcommand{\lambdai}{\lambda_{_{\mathrm{i}}}}

\newcommand{\qbar}{\overline{q}}
\newcommand{\qm}{q_{\mathrm{m}}}

\newcommand{\qprime}{\dot{q}}
\begin{document}
\title{Optimal work of Brownian motion in a harmonic time-dependent stiffness potential. Effect of the initial position.}
\author{Pedro J. Colmenares}
\affiliation{Universidad de Los Andes. Departamento de Qu\'{\i}mica . M\'erida 5101 - Venezuela}
\email{gochocol@gmail.com}
\thanks{Corresponding Author}
\author{Oscar Paredes-Altuve}
\email{paredes.oscar@gmail.com}
\affiliation{Universidad de Chile. Facultad de Ciencias F\'{\i}sicas y Matem\'aticas.  Departamento de F\'{\i}sica. Santiago - Chile.}

\begin{abstract}
The system consists of a Brownian particle immersed in a heat bath trapped in optical tweezers with a time-dependent strength acting as an external protocol. In [Phys. Rev. Letts., 98:108301, 2007] the optimal mean work in the overdamped regime was thoroughly calculated by assuming the work must be averaged over the distribution of the initial position of the particle. The present research assumes instead the solution of the Langevin equation for any given initial position and its average done over the noise distribution. Therefore, this proposal extends in a more general sense the results already published, including the appearance of Maxwell's demon for particular initial conditions which is analyzed in terms of entropy production rate and the mutual information obtained by measuring the particle position. The proposed research has the advantage of being able to be compared with data from numerical simulations.
\end{abstract}
 \pacs{05.30.?d; 05.40.Jc}
\keywords{Brownian motion, Stochastic processes.}
\maketitle

\section{Introduction}
\label{Sec0}

A  Brownian particle immersed in a heat bath at a given temperature and interacting with a  time-dependent harmonic potential is an iconic small system analyzed under the framework of stochastic energetics \cite{Sekimoto,SeifertReport}. In particular, when the bath is at the temperature $T$ and the external field is  a time-dependent harmonic potential  $V(q,\tau)=\lambda(\tau)q^{2}/2$ where $q$ is the fixed position of the particle and the strength $\lambda(\tau)$ is an external protocol, the mechanical work is a functional of the driving given by \cite{SchmiedlSeifert2007,SeifertReport},

\bea
W[\lambda(\tau)]&=&\int_{0}^{t}d\tau\,\dot{\lambda}(\tau)\left<\frac{\partial}{\partial\lambda(\tau)}V(q(\tau),\lambda(\tau)\right>,\nonumber\\
&=&\frac{1}{2}\int_{0}^{t}d\tau\,\dot{\lambda}(\tau)\left<q^{2}(\tau)\right>,
\label{wave}
\eea
assuming the system is at equilibrium when the protocol $\lambda(t)$ is turned on at $t=0$. The dot on a variable denotes its time derivative.

The particle position is a functional of the protocol, so to find the optimal mean work it must be optimized in terms of $\lambda(\tau)$. The latter is a smooth function of time that must be determined by variational methods such that in a finite time the work induced by the external field be minimal. This was thoroughly investigated by Schmiedl and Seifert \cite{SchmiedlSeifert2007} for the overdamped Langevin equation assuming the average in Eq. (\ref{wave}) over the noise distribution is also applicable for a general initial position drawn from its initial equilibrium distribution. Other lines of research aim toward a different objective, such as for instance, the determination of the work probability density function considering its results due to bath noise. This was addressed by Chvosta {\it et al.} \cite{ChvostaEtAl} for piecewise constant protocols. Both investigations were based on dynamics where the inertial effects of the particle are negligible.

Our central objective is to determine for an overdamped particle positioned in any initial position, the optimal protocol and the mean work such that the average should be performed only over all possible outcomes of the thermal noise. For this, we will make use of a variational procedure to determine the optimal protocol originally designed in \cite{SchmiedlSeifert2007}. We will show that the results obtained with the latter are a particular solution of our broader scheme that involves more general initial conditions. Furthermore, the results obtained could be verified with simulation data.

The manuscript is organized as follows. First, we derive the expression for the optimal work in terms of the solution of the overdamped Langevin equation to get the required dependence on the initial position. There follows then an analysis of the consequences that an experimental position measurement has on the real position distribution. This is done with the purpose to determine the effect of the initial position on the amount of information obtained in the measurement.  The results of the preceding sections are discussed next emphasizing the reproduction of previous findings and the explanation of the unexpected appearance of Maxwell's demon in terms of entropy production rate and mutual information. The article ends with a summary including the equivalence of the variational method with the so-called ``shortcut of isothermality'' of Li {\it et al.} \cite{LiQuanTu}.

\section{General equations}
\label{Sec1}
We suppose the dynamics follow the overdamped Langevin equation
\be
\qprime(\tau)=-\gamma\,\lambda(\tau)\,q(\tau)+\sqrt{2\,\kb\gamma\,T}\,\xi(\tau),
\label{qov}
\ee
where $\gamma$ is the friction coefficient of the thermal bath and $\kb$ is the Boltzmann constant. The noise $\xi(\tau)$ is a zero-mean and delta correlated white noise $\left<\xi(\tau)\,\xi(s)\right>=\delta(\tau-s)$. It will be assumed for simplicity that $\{\gamma,\kb,T\}$  are set to one. 

 The solution of the  Langevin equation functionally depends on $\lambda(\tau)$. 
It reads as:

\bea
q(\tau)&=&\qbar(\tau)+\varphi(\tau),\label{lepj}\\
\qbar(\tau)&=&\qzero\,\eul^{-h(\tau)},\label{qbar}\\
\varphi(\tau)&=&e^{-h(\tau)}\int_{0}^{\tau}dt\,\eul^{h(t)}\,\xi(t),\\
h(\tau)&=&\int_{0}^{\tau}dt\,\lambda(t),
\eea
where $\qzero$ is the initial position and the new noise $\varphi(\tau)$ is a colored Gaussian zero-mean with a correlation function also depending on the driving and given by \cite{Gardiner1}
\be
\left<\varphi(\tau)\varphi(s)\right>=2\,\eul^{-h(\tau)-h(s)}\int_{0}^{\mbox{\tiny{Min}}\{\tau,s\}}dt\,\mbox{e}^{2h(t)}.
\label{corrnoise}
\ee

Let $u(\tau)=\left<q^{2}(\tau)\right>$ where the average is over the noise density. Then,
\bea
u(\tau)&=&\qzero^{2}\,\eul^{-2h(\tau)}+2\,\mbox{e}^{-2\,h(\tau)}\int_{0}^{\tau}dt\,\eul^{2\,h(t)}.\label{ut}
\eea

To complete the description, the probability density $p(q,\tau |\qzero)$ associated with Eq. (\ref{qov}) satisfies a Fokker-Planck equation which according to Ito's formula \cite{Gardiner1} is, 
\bea
\frac{\partial p(q,\tau)}{\partial t}&=&\lambda(\tau)\frac{\partial}{\partial q}q\,p(q,\tau)+\frac{\partial^{2}p(q,\tau)}{\partial q^{2}}.
\label{FPE}
\eea

The solution of this equation is a Gaussian corresponding to that of an Ornstein-Uhlenbeck process with mean $\qbar(\tau)$ and standard deviation $\sigma^{2}(\tau)=u(\tau)-\qbar^{2}(\tau)$ \cite{Gardiner1,Risken}. The stationary solution (ss) is  respectively, $p^{\mathrm{ss}}(q,\tau)=\sqrt{\lambda(\tau)/\pi}\exp[-\lambda(\tau)\,q^{2}/2]$.

Likewise, the mean work in the whole time interval is obtained from  the integration of Eq. (\ref{wave}) \cite{SchmiedlSeifert2007}
\be
W[\lambda(\tau)]=\frac{1}{2}\bigg[\lambda(\tau)\,u(\tau)-\ln u(\tau)\bigg]_{0}^{t}+\frac{1}{4}\int_{0}^{t}\!\!d\tau\frac{\dot{u}(\tau)}{u(\tau)}.
\label{meanwork1}
\ee

Because it depends non-locally on $\lambda(\tau)$ through $u(\tau)$, its optimization leads to tangled equations whose solution requires demanding procedures, which are mostly numerical. Instead, we apply the method  of \cite{SchmiedlSeifert2007} based on casting the functional as a simple equation depending locally on $u(\tau)$ and  of its first time-derivatives, i.e,
\be
 \dot{u}(\tau)+2\,\lambda(\tau)\,u(\tau)=2, 
 \label{pj2}
 \ee
 leading to the Euler -Lagrange equation $\dot{u}^{2}(\tau)-2\,\ddot{u}(\tau)\,u(\tau)=0$  whose solution renders \cite{Math} 
\be
u(\tau)=C_{3}\,\tau\left(1+\frac{C_{3}\,\tau}{4\,C_{4}}\right)+C_{4}.
\label{pj}
\ee

The final result depends upon the identification of the constants. 
In \cite{SchmiedlSeifert2007} a particular initial position, drawn at random from the equilibrium steady
state of the system, was considered. The steady value is  $u(0)=C_{4}=1/\lambda_{i}$ with $\lambdai$ being the pre-set initial protocol value. 
Rather, we want to analyze the explicit dependence of the dynamics in terms of arbitrary initial positions, which in turn, are relevant in a comparison with molecular dynamics simulation data. This approach then requires an initial $u(0)=\qzero^{2}$ as can be seen from Eq. (\ref{ut}). Thus, our results extend the aforementioned findings to any values of $\qzero$ generating new predictions on the outcomes of the functions to be determined, in particular, the show-up of Maxwell's demons. The replication of the results shown in \cite{SchmiedlSeifert2007} is included as it should be.

Accordingly, then $C_{4}=\qzero^{2}$ and $C_{3}$ is determined by minimizing Ref.{meanwork1} with respect to this constant. It reads
\be
C_{3}=\frac{-2\,\qzero^{2}\left(1+t\,\lambdaf\right)+2\,\qzero\sqrt{\qzero^{2}+t\,\left(2+t\,\lambdaf\right)}}{t\,\left(2+t\,\lambdaf\right)},
\ee
where $\lambdaf$ is the pre-fixed value of the driving at the end of its application.

The optimal protocol is found from Eq. (\ref{pj2}). It reads,
\be
\lambda(\tau)=\frac{2\,\left(2-C_{3}\right)\qzero^{2}-C_{3}^{2}\,\tau}{\left(C_{3}\,\tau+2\,\qzero^{2}\right)^{2}},
\label{lambdat}
\ee
valid for $0^{+}<\tau<t^{-}$.  As will be seen below, the protocol also implies jumps at the beginning and also at the end.

The mean work in the whole time interval is obtained from  Eq. (\ref{meanwork1}) giving

\bea
&W&\!\!\!(t)=\frac{1}{8\,\qzero^{2}}\Bigg[C_{3}\,t\left[4\,\qzero^{2}\lambdaf+C_{3}\,\left(2+\lambdaf\,t\right)\right]\nonumber\\
&+\!\!\!\!&4\,\qzero^{2}\left[\qzero^{2}\,(\lambdaf\!-\!\lambdai)\!-\!2\ln\left(2\,(C_{3}\,t\!+\!2\,\qzero^{2})\right)\right]\Bigg].
\label{meanwork}
\eea

In an experiment where the optical trap is turned on and off almost instantaneously, there is neither exchange of heat with the reservoir nor changes in the position and velocity of the particle. The instantaneous adiabatic work is that which is involved. It is obtained from the work equation of \cite{SchmiedlSeifert2007} by making the integral term zero and subsequently evaluating the expression at $t=0$. The result is
\be
W_{\mathrm{ins}}= \frac{1}{2}\,\qzero^{2}\,(\lambdaf-\lambdai),
\label{inst}
\ee
which will be used below.

Equations Ref. {lambdat} and Eq. (\ref {meanwork}) will be evaluated in the third section.

\section{Mutual information}
\label{Sec1b}

The theory should be confronted with experimental data to quantify the accuracy of its theoretical prediction. There has already been developed a procedure to determine the distribution of the position conditioned to its experimental measurement of an overdamped Brownian particle in a moving harmonic potential interacting with a heat bath \cite{AbreuSeifert}. We used this to determine if our equations depend rather on the initial position of the particle than on its average, corresponding to the center of the potential as in \cite{AbreuSeifert}.

The distribution of the particle position is
\be
p(q,\tau\mid\qzero)=\frac{1}{\sqrt{2\,\pi\,\sigma^{2}(\tau)}}\exp\!\!\bigg[-\frac{[q-\qzero\,\Delta(\tau)]^{2}}{2\,\sigma^{2}(\tau)}\bigg], 
\label{pqq0pj}
\ee
with $\Delta(\tau)=\exp[-h(\tau)]$. Likewise, the initial dis\-trib\-ution is a delta function centered at a given $\qzero=\alpha$, that is, $p(\qzero)=\delta(\qzero-\alpha)$.

The experimental setup is supposed should be designed such that the trajectory of the particle is measured. Moreover, we consider that each measurement $\qm$ of the real position $q$ with precision $\epsilon$ is instantaneous. The distribution of $\qm$ about around $q$ is 
\be
p(\qm,\tau|\,q)=\frac{1}{\sqrt{2\,\pi\,\epsilon^{2}}}\exp\bigg[-\frac{(\qm-q)^{2}}{2\epsilon^{2}}\bigg],
\ee
where $q$ is the position $q(\tau)$ at the time of the measurement.

The marginal distribution $p(q,\tau)$ is obtained from Eqs. (\ref{pqq0pj}) and $p(\qzero)$ as 

\bea
p(q,\tau)&=&\int_{-\infty}^{\infty}d\qzero\,p(q,\tau |\,\qzero)\,p(\qzero),\nonumber\\
&=&\frac{1}{\sqrt{2\,\pi\sigma^{2}(\tau)}}\exp\!\bigg[-\frac{(q-\alpha\,\Delta(\tau))^{2}}{2\,\sigma^{2}(\tau)}\bigg].
\eea

Likewise, $p(\qm,\tau)$ is given by:

\bea
p(\!\!\!\!\!&\qm&\!\!\!\!\!,\tau)=\!\!\int_{-\infty}^{\infty} dq\,p(\qm\tau |\,q)\,p(q,\tau),\nonumber\\
&=&\frac{1}{\sqrt{2\pi(\epsilon^{2}+\sigma^{2}(\tau))}}\exp\!\bigg[\frac{(\qm-\alpha\,\Delta(\tau))^{2}}{2\,(\epsilon^{2}+{\sigma^{2}(\tau))}}\bigg].
\eea

According to Bayes's theorem 
\be
p(q,\tau| \,\qm)\,p(\qm)=p(\qm,\tau| \,q)\,p(q),
\ee
therefore
\bea 
p(q,\tau|\,\qm)_{\mathrm{q}_{_{0}}}\!\!\!\!\!\!&=&\!\!\!\!\!\!\frac{1}{\sqrt{2\,\pi\,\sigma_{\mathrm{m}}^{2}(\tau)}}\nonumber\\
&\times&\!\!\!\!\exp\bigg[-\frac{\left(q-\qm\,\kappa_{_{1}}(\tau)-\kappa_{_{2}}(\tau)\right)^{2}}{2\,\sigma_{\mathrm{m}}^{2}(\tau)}\bigg],\label{pqmqpj}
\eea
where the subindex $\mathrm{q}_{_{0}}$ was added to take into account the implicit dependence on the particle's initial position since $\sigma^{2}(\tau)$ is a functional of the optimal protocol which in turn depends on $\qzero$. Furthermore, $\kappa_{_{1}}(\tau)=\sigma^{2}(\tau)/\left(\epsilon^{2}+\sigma^{2}(\tau)\right)$, $\kappa_{_{2}}(\tau)=\alpha\,\epsilon^{2}\,\Delta(\tau)/\left(\epsilon^{2}+\sigma^{2}(\tau)\right)$ and $\sigma_{\mathrm{m}}^{2}(\tau)=\epsilon^{2}\kappa_{_{1}}(\tau)$, respectively.

An important aspect of this result is the quantification of the distinguishability of the $q$ and $\qm$ distributions each time measurement of the particle position is done. It is given by the Kullback-Leibler distance \cite{CoverThomas}

\bea
I(\qm,\tau)&=&\int_{-\infty}^{\infty}dq\,p(q,\tau|\,\qm)_{\mathrm{q}_{_{0}}}\!\!\ln\bigg[\frac{p(q,\tau|\,\qm)_{\mathrm{q}_{_{0}}}}{p(q,\tau)}\bigg],\nonumber\\
&=&\frac{\sigma^{2}(\tau)}{2(\epsilon^{2}+\sigma^{2}(\tau))^{2}}\left((\qm\!-\!\alpha\,\Delta(\tau))^{2}\!-\!1\right)\nonumber\\
&-&\!\frac{1}{2}\ln\left[\frac{\epsilon^{2}\,\kappa_{_{1}}(\tau)}{\sigma^{2}(\tau)}\right],
\eea
equivalent to the information gained from the measurement. Its integration over $p(\qm,\tau)$ is the so-called mutual information or relative entropy, i.e,

\bea
\mathcal{I}(\tau)&=&\int_{-\infty}^{\infty}d\qm\,p(\qm,\tau)\,I(\qm,\tau),\nonumber\\
&=&\frac{1}{2}\ln\left[1+\frac{\sigma^{2}(\tau)}{\epsilon^{2}}\right],
\label{mutual}
\eea
measuring how distinct the two distributions are and being a useful relation between dissipation and irreversibility, as well \cite{SeifertReport}. 

Surprisingly, this result has the same functional form as the one obtained in  \cite{AbreuSeifert} for the moving harmonic potential, even though the derivations started from different probability densities. This allows us to conclude that no matter how the harmonic potential depends on time, the mutual information retains its functional form. It makes sense since the measurement is instantaneous in the two processes, the external agent will always perform it on a particular parabolic-kind potential acting in the background. The differences are reflected in the standard deviation of the distribution of the actual position of the particle. 

Using Eq. \ref{pqq0pj}, the total entropy 
\bea
S(t)&=&-\int_{-\infty}^{\infty}dq\,p(q,\tau|\,\qzero)\ln\left(p(q,\tau|\,\qzero)\right),\nonumber\\
&=&\frac{1}{2}\left(1+\ln(2\,\pi\,\sigma^{2}(\tau)\right),
\eea
which differs from  $\mathcal{I}(\tau)$ by a constant related to the precision of the measurement.  
\section{Discussion of results}
\label{Sec2}

We proceed to make the calculations taking into account that they depend on the initial position $\qzero$. It will be assumed $\lambdaf \!>\! \lambdai$  to add the requirement that the potential is wider compared to the initial one. The calculation of the mean work is modulated by $\lambdai$. This is not the case with the protocol because it is independent of it, so any value of it could be chosen. However, it is not admissible as its value is restricted by the aforementioned condition.

\begin{figure}[ht]
\includegraphics[height=13cm,width=10cm]{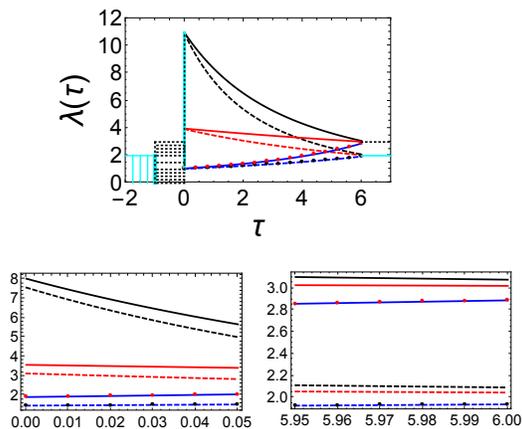}
\vspace{-6.5cm}
\caption{ Effect of $\qzero$ and $\lambdaf$ on the optimal protocol $\lambda(\tau)$ for $\qzero=0.3$ (black), $\qzero=0.5$ (red) and $\qzero=1$ (blue). Solid curves are for $\lambdaf=2$ and dashed ones for $\lambdaf=3$. The bottom graphics expand the area right at the beginning and final protocol application time  $\tau$ ; $\lambdai=1$. The blue and red dots are the solutions of \cite{SchmiedlSeifert2007}. The boxes at the left show the range of the allowed $\lambdai$.}
\label{Fig1}
\end{figure}

The optimal protocol is shown in Fig. \ref{Fig1}. The curves were determined for $\qzero$ of 0.3 (black), 0.5 (red), 1 (blue), and $\lambdai=1$. Jumps at the beginning and the end with different magnitudes are observed. The jumps at the start are the same regardless of the chosen $\lambdai$ and are shown at the bottom for the area at the beginning and end of the protocol. The $\lambdai$ allowed by the condition $\lambdaf>\lambdai$  are represented by the two boxes at the left of the figure. The solid curves are for $\lambdaf=2$ and the dashed ones for $\lambdaf=3$. The red and blue dot curves are the results obtained in \cite{SchmiedlSeifert2007} for $\lambdai=\qzero=1$ and $\lambdaf$ of 2 and 3, respectively. The concordance observed between our results and those of  \cite{SchmiedlSeifert2007} is because the former $u(0)=1$. It should be noted that this is one among many without invalidating the existence of the latter. As long as $\qzero=1/\sqrt{\lambdai}$ our results will agree with those of \cite{SchmiedlSeifert2007}. In other words, their findings are particular results of this proposal. In this sense, the scheme presented here is valid for any $\qzero$.
\begin{figure}[h]
\centering
\includegraphics[height=13cm,width=9.5cm]{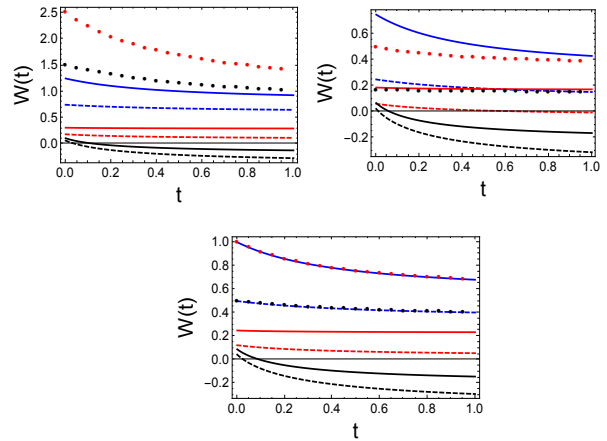}
\vspace{-7cm}
\caption{ Effect of $\qzero$ and $\lambdaf$ on the mean work $W(t)$ for the same $\{\qzero,\lambdaf\}$ conditions  of Ref.{Fig1}. The left top panel is for $\lambdai=0.5$ and the right one for 1.5 while the bottom is or $\lambdai=1.0$. The red and blue dots are the solutions from \cite{SchmiedlSeifert2007}.}
\label{Fig2}
\end{figure}

The initial $\lambdai$ has a bearing on the mean work. This is shown in the two panels of Fig. \ref{Fig2} for values of 0.5 (left top), 1.5 (right-top), and 1.0 (bottom) and the protocols depicted in Fig. \ref{Fig1}. We observe in the three chosen conditions the particle doing average work against the external field with different intensities for sufficiently low values of $\qzero$ such as is the case for 0.3 (black curves). Later, the emergence of this attribute will be explained by other points of view. As expected, the bottom panel shows the agreement mentioned before.

To obtain the version equivalent to figure 1a of \cite{SchmiedlSeifert2007} it is necessary to write first Eq. \ref{meanwork} in the same reduced variable system. Defining $x=\tau/t$, $y=\lambdaf/\lambdai$,  $z=\lambdai t$ and the extra parameter $w=\qzero^{2}\lambdai$ we get a closed expression of $\lambda(x,z)/\lambdai$ parametrized by $y$ and $w$, that is,

\bea
\frac{\lambda(x,z)}{\lambdai}\!\!\!\!\!&=&\!\!\!\!\!\frac{w}{A_{1}(x,z)}\!\!\left[-w \left(x\! \left(y^2 z^2\!+\!2\, y\, z\!+\!2\right)\!-\!y^{2}\,z^{2}\right.\right.\nonumber\\
&-&\left. 3\, y \,z\!-\!2\right)-\alpha_{_{1}}(z) \, (2\, x\, (y  \, z\!+\!1)\!-\!y\, z\!-\!2)\nonumber\\
&+&\left. z\, (\,y \,z+2) (x-y\, z-2)\right], \\ 
 A_{1}(x,z)&=&z \,(w\, (x\, y\, z+x-y\, z-2)-\alpha_{_{1}}  x)^2,  \\
 \hspace{0.5cm}\alpha_{_{1}}(z)&=&\sqrt{w\, (w+z\, (y\, z+2))},
\eea

\begin{figure}[ht]
\centering
\includegraphics[height=13cm,width=10cm]{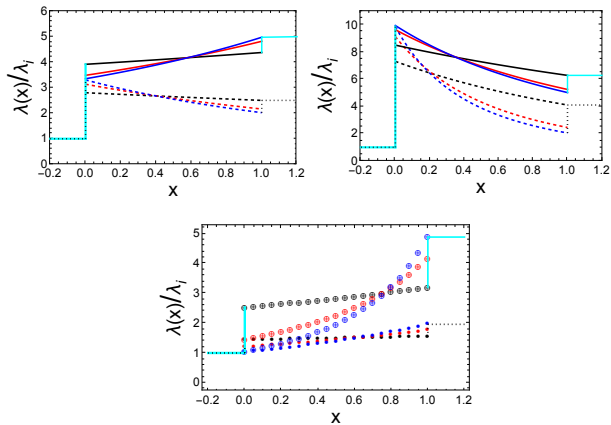}
\vspace{-6.5cm}
\caption{Optimal protocol $\lambda(x)/\lambdai$ as a function of the scaled time $x$ for the set of $\{y,z\}$ of \cite{SchmiedlSeifert2007} and the extra parameter $w$ of 0.3 ( top left ) and 0.1 (top right). The bottom plot shows the corresponding for $w=1$ as in \cite{SchmiedlSeifert2007}. See text for more details.}
\label{Fig3}
\end{figure}

The function $\lambda(x)/\lambdai$ is plotted in Fig. \ref{Fig3} where the left graphic is for $w=0.3$ while the right is for $w=0.1$.
The $\{y,z\}$ parameter sets resemble those of \cite{SchmiedlSeifert2007}.  They are identified as dashed curves for  $\{2,0.1\}$ (black), $\{2,1\}$ (red), and $\{2,10\}$ (blue) while the solid ones are for the remaining of $\{5,0.1\}$, $\{5,1\}$  and $\{5,10\}$ with the same color denomination.  The magnitudes of the jumps at the beginning and the end are represented as cyan and black dotted lines. The particular case for  $w=1$ case (bottom plot) corresponds to that of \cite{SchmiedlSeifert2007}. It is important to remark that it can also be obtained if the initial condition $u(0)=1/\lambdai$ is replaced in Eq. (\ref{pj}) with $\qzero=\lambdai=1$ and $\lambdaf=\{2,3\}$.

The figure analogous to 1b of \cite{SchmiedlSeifert2007} requires the derivation of the mean work, Eq. Ref.{meanwork}, in terms of $z$ and to be parameterized by $y$ and $w$, i.e.,

\bea
W(z)&=&-\frac{1}{A_{2}(z)}\left[w \left(2\, \alpha_{_{2}}(z) +y \,z^{2}-2 \,y \,z+2\, z-2\right)\right.\nonumber\\
&+&\left. 2\,z \,(y\, z+2)\log\left( \alpha_{_{3}}(z)\right) -z\, (y\, z+2)\right], \\
 A_{2}(z)&=&2\, z\, (y\, z+2),  \\
 \alpha_{_{2}}(z)&=&\sqrt{\frac{w+z\, (y\, z+2)}{w}},  \\
 \alpha_{_{3}}(z)&=&\frac{\alpha_{_{2}}(z)  +1}{y\, z+2}.
\eea

\begin{figure}[h]
\centering
\includegraphics[height=12cm,width=8cm]{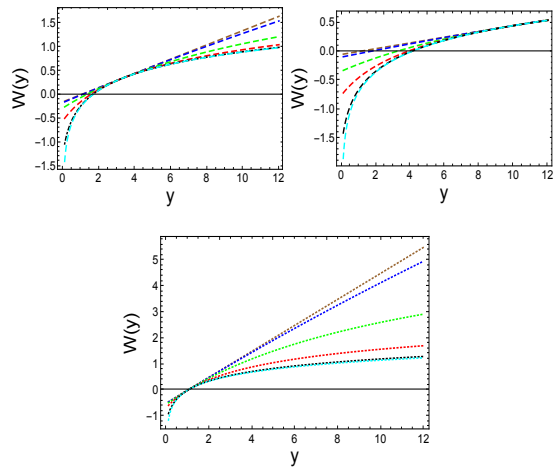}
\vspace{-5.cm}
\caption{ Optimal mean work as function of  $y$ for the same set of $z$ parameters of figure 1b of \cite{SchmiedlSeifert2007}. The left graphic at the top is for $w$ of 0.3 and the right is for $w=0.1$. The lower plot is for $w=1$ as in \cite{SchmiedlSeifert2007} but extended for $y<1$. See the text for the identification of the dashed and dotted curves.}
\label{Fig4}
\end{figure}

This figure (analogous to figure 1b of Ref. \cite{SchmiedlSeifert2007}) is presented as Fig. \ref{Fig4}. The $z$ values are those of the inset of figure 1b of \cite{SchmiedlSeifert2007}, identified as 0 (brown), 0.02 (blue), 0.2 (green), 1 (red), 10 (black), and $\infty$ (cyan), respectively.
The curves were calculated for $w=0.3$ (top left) and $w=0.1$ at the right. For $w=1$, the resulting plot is shown at the bottom and is just figure 1b of \cite{SchmiedlSeifert2007} but extended for $y<1$. The appearance of negative work in this figure along with the ones of Fig. \ref{Fig2}, not shown in \cite{SchmiedlSeifert2007}, forces us to carry out a more detailed search for other parameters of the system. It is noticed an increasing negative work for a decreasing  $w$ is in agreement with the results shown in Fig. \ref{Fig2}. The initial ($z\rightarrow 0$) and quasi-static ($z\rightarrow\infty$) mean works are given by:
\bea
W(0)&=&\frac{w}{2}\,(y-1),\label{W0}\\
W(\infty)&=&\frac{1}{2}\,\left(1+\ln (w\,y)-w\right)\label{West},
\eea
where they match those of \cite{SchmiedlSeifert2007} replacing $w=1$. $W(0)$ is also the adiabatic $W_{\mathrm{ins}} $ given by Ref.{inst} in reduce variables.

\begin{figure}[ht]
\centering
\includegraphics[height=12cm,width=9.5cm]{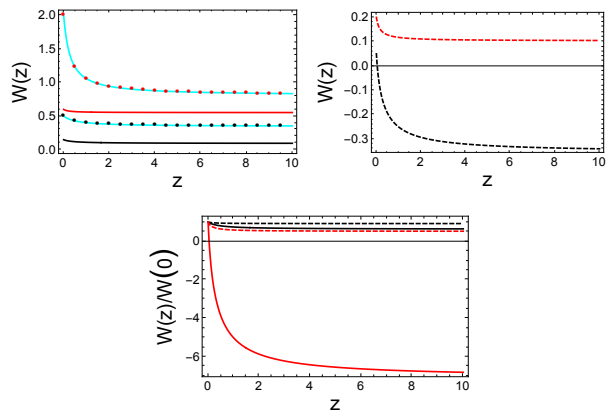}
\vspace{-6cm}
\caption{ Optimal mean work as function of  $z$ for $w=0.3$ (top left) and $w=0.1$ (top right). The values of $y$ are 2 (black) and 5 (red). The bottom shows the combination of the top plots. }
\label{Fig5}
\end{figure}

$W (z)$ is plotted versus $z$ in Fig. \ref{Fig5} to check any occurrence of $ W (z) < 0$. The black and red dotted curves are the solutions obtained from \cite{SchmiedlSeifert2007} for $w=1$. They superimpose the cyan curves of this article. Solid curves are for $w=0.3$ and 0.1 for the dashed ones. The parameter $y=2$ is represented by black curves while $y=5$ by red. Here, $W(z)<0$ is obtained for low values of $w$ and $y$.
The ratio $W(z)/W(0)$ is shown at the bottom. As before, negative work is observed for a low $y$ value. In particular, the occurrence of negative work is mainly maintained throughout $z$  being positive for a short $z$ interval around zero.

It is a physical fact that the larger $\qzero$, the wider the potential. Thus, when $\qzero$ is such that the particles do work, they are more tightly packed compared to when the work is done by the field. The action of the field tends to heat them and since the process is isothermal, they do work against the field as well as transfer heat to the bath to lower the temperature. For larger values of the initial position, the packing decreases, the field does the work, and excess heat is transferred to the reservoir. This phenomenon also has its strictly mathematical explanation through Landauer's erasure principle \cite{Landauer} applied to a particular Maxwell's demon (see below) that collects specific information about the state of the system which is later transformed into work. For a further revision of the erasure procedure and the role of information in quantum-information theory see the review by Maruyama {\it et al.} \cite{MaruyamaNoriVedral} and references therein. Recently, Parrondo   {\it et al.} \cite{ParrondoEtAl} discussed the thermodynamics of memory, feedback processes, and information flows from the perspective of stochastic processes for a system in contact with a single thermal bath.
\begin{figure}[ht]
\centering
\includegraphics[height=12cm,width=9cm]{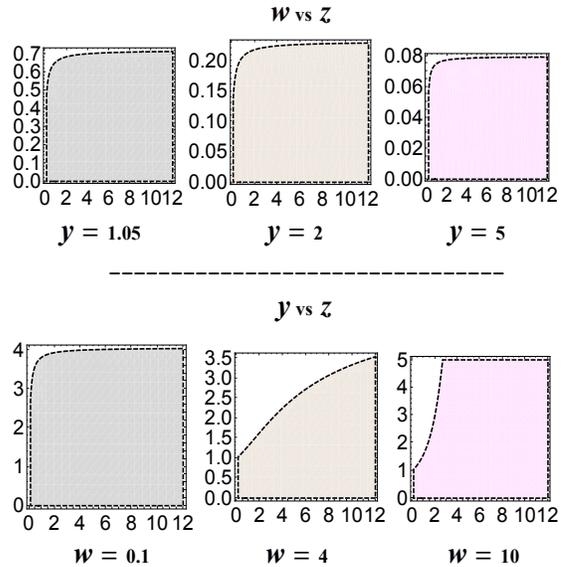}
\vspace{-4cm}
\caption{Set of $w$, $y$ and $z$ which makes $W<0$. }
\label{Fig6}
\end{figure}

The generalization of the last two figures can be obtained by finding the set $\{w,y,z\}$ which makes $W<0$. This is shown in Fig. \ref{Fig6}. The case $w=1$ (not shown) gives $W<0$ for $y\in\{0,1\}$, which was not considered in \cite{SchmiedlSeifert2007}, and is clearly seen in Fig. \ref{Fig4}.

It can be inferred then that this new set of results could in principle be tested in molecular dynamics simulations where results depend upon the particle's initial position. 

 \begin{figure}[ht]
\centering
\includegraphics[height=14cm,width=11cm]{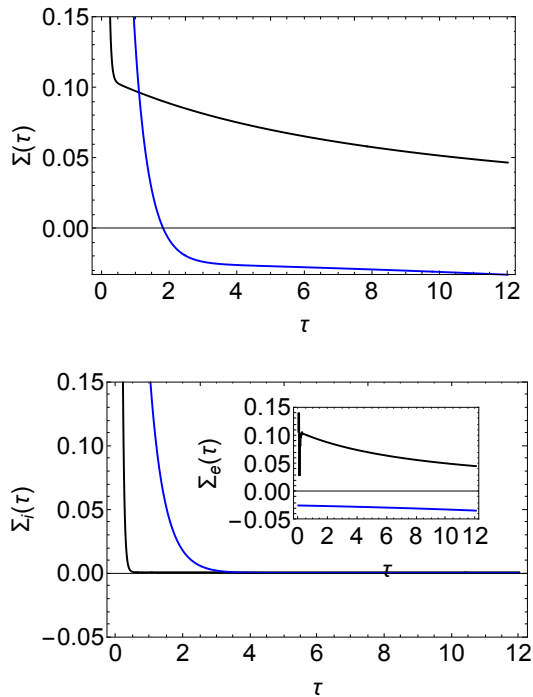}
\vspace{-3.45cm}
\caption{ EPR total $\Sigma(\tau)$, internal $\Sigma_{i}(\tau)$ and flow to the reservoir $\Sigma_{e}(\tau)$ for the black and blue solid curves of Fig. \ref{Fig2}. }
\label{Fig7}
\end{figure}
 \begin{figure}[ht]
\centering
\includegraphics[height=13cm,width=11cm]{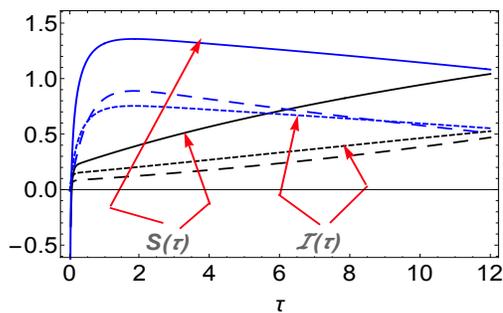}
\vspace{-8.cm}
\caption{ Mutual information and total entropy of the system corresponding to Fig. \ref{Fig8} with their associated standard deviations shown as long-dashed curves.}
\label{Fig8}
\end{figure}
In a system in equilibrium, the correlation of the fluctuations for small external disturbances is given in terms of the response function. This is not the case in stationary systems out of equilibrium because the detailed balance relation is broken and therefore there is a continuous degradation of energy to the thermal reservoir \cite{Seifert,BroeckEsposito,RoldanEtAl}. Since we are interested in the production of work, the associated entropy production rate (EPR) can be determined from the time derivative of Gibbs entropy. Defining the probability flux of the FPE, Eq, (\ref{FPE}), by $J(q,t)=-q\, p(q,t)-\partial p(q,t)/\partial q$ with $p(q,t|\,\qzero)$ given by the Gaussian Eq. \ref{pqq0pj}, the corresponding total EPRs, namely, the total $\Sigma(\tau)$, the internal due to the constituting non-equilibrium process along the dynamics $\Sigma_{i}(\tau)$ and the flow into the reservoir $\Sigma_{e}(\tau)$ are \cite{BroeckEsposito}:
\bea
\Sigma(\tau)&=&-\int_{-\infty}^{\infty}dq\,\dot{p}(q,\tau)\ln\,p(q,\tau),\nonumber\\
&=&\frac{1}{\sigma^{2}(\tau)}-\lambda(\tau),\\
\Sigma_{i}(\tau)&=&\int_{-\infty}^{\infty}dq \,\frac{J^{2}(q,\tau)}{p(q,\tau)},\nonumber\\
&=&\frac{1}{\sigma^{2}(\tau)}\!-\!2\lambda(\tau)\!+\!\lambda^{2}(\tau)\left(\qbar^{2}(\tau)+\sigma^{2}(\tau)\right)\!\!,\\
\Sigma_{e}(\tau)&=&-\int_{-\infty}^{\infty}\!\!dq\,J(q,\tau)\,J_{_{\!\lambda}}(q,\tau),\nonumber\\
&=&\lambda(\tau)-\lambda^{2}(\tau)\left(\qbar^{2}(\tau)+\sigma^{2}(\tau)\right),
\eea
where $J_{_{\!\lambda}}(q,\tau)=-q\,\lambda(t)$. Choosing the black and blue curves of Fig. \ref{Fig2} as prototypes of negative and positive works with parameters $\{\qzero\lambdai,\lambdaf\}$ of $\{0.3,1,2\}$ and $\{1,3,3\}$, respectively, the total EPR and its components are shown in Fig. \ref{Fig7}. Notice the Second Law is preserved since $\Sigma_{i}(\tau)>0$.  Regardless of the work sign, there is always a degradation of field energy toward the reservoir. However, to preserve the total balance, the transformation of the information into work acquired by the "Maxwell creature" does not produce any EPR at all, leaving as a contribution to the balance that part due to the heat transferred to the thermal bath. In the other case, the largest contribution to the total EPR comes from the heat dissipated to the bath which in turn is greater than the contribution of $\Sigma_{i}(\tau)$. As expected, the entropy production $\Sigma_{i}(\tau)$ is higher when the work is done by the field. The standard deviations used in the calculation are shown in Fig. \ref{Fig8} as the long-dashed curves.

Finally, the mutual information and statistical entropy are shown in Fig. \ref{Fig8}. The first is modulated by $\epsilon^{2}$ with the two functions exhibiting the same topology of the associated standard deviation shown as a long dashed curve. The steady-state behavior of the mean work at large times is captured by both properties regardless of the initial position. As expected, the inequality $\cal{I}(\tau)\leqslant S(\tau)$ holds \cite{SagawaUeda1}. From the figure, it is found that whenever the work is done by the particle, the experimental measurement of the actual position leads to a decrease in both the total and relative entropy, compared to when the work is done by the external field. 

In the narrative of Maxwell's demons, the ``creature'' would be categorized as of the ``smart'' kind whose first function would be to measure the mean initial position of the entire system and select the associated position distribution that leads to work extraction. Once this is achieved, then operates like an ordinary demon by selecting the fast particles from the slow ones and placing them separately on each side of the virtual gate. Finally, it transfers heat from the ``fast'' side to the ``slow'' to generate work against the external agent.
Regarding the previous thermodynamic analysis, the information collected in the process is the internal contribution to the total Shannon's entropy which after its erasure gives the expected negative work.

\section{Final remarks}
\label{Sec3}

It has been shown that new and important information about the thermodynamics of the system can be obtained by choosing the appropriate initial conditions of the dynamics. In particular, the occurrence of negative mean works depends upon these conditions. This leads us to consider that the obtained results could potentially be used in comparison with molecular simulation data. 

The experiments by Kahn and Sood \cite{KhanSood} on colloidal particles in equilibrium trapped in high-intensity optical tweezers could be the starting point for designing a convenient experimental set-up that allows the irreversible work to be measured through the Jarzynski relation \cite{Jarzynski1}.

There is a close connection between the variational procedure  \cite{SchmiedlSeifert2007} used in this proposal to determine the optimal protocol and the so-called ``{\it shortcut to isothermality}'' approach of Li {\it et al.} \cite{LiQuanTu} where the potential is modified in such a way that the dynamic remains isothermal and equilibrium is preserved in the initial and final states. According to \cite{LiQuanTu}, the external potential must be modified to $V((q,\tau) =\lambda_{\mathrm{eff}}(\tau)\,q^{2}/2$ with $\lambda_ {\mathrm{eff}}(\tau)=\lambda(\tau)+(1/2)\dot{\lambda}(\tau)/\,\lambda(\tau)$ with the additional constraint that $\dot{\lambda}(\tau)$ must vanish at the beginning and end of the protocol. The role of the additional term in the new potential is to allow a quick evolution into target states of otherwise slow isothermal dynamics preserving the volume of the phase space. The strategy is based on borrowing a similar idea originally developed for quantum and classical adiabatic systems and thoroughly analyzed by several authors as seen in the review by Gu\'ery-Odelin {\it et al.} \cite{ReviewSTA}. If the optimal protocol derived above is used in $\lambda_{\mathrm{eff}}$, the resulting effective driving overlaps the curves presented in Fig. \ref{Fig1}. This demonstrates without any margin of doubt that the initial and final jumps implicit in the expression of $\lambda(\tau)$ are a guarantee that the system will remain in the same equilibrium state at the beginning and end of the driving. In addition, the temperature will be constant throughout the dynamics.

The solution for systems out of the overdamped deserves to be investigated. It will allow the analysis of the behavior of the mean work for low friction coefficient values of the heat reservoir.

\section*{Acknowledments}
The authors thank Prof. Roy Little and Cesare Olinto Colasante of Universidad de Los Andes for their useful suggestions. 

\bibliographystyle{apsrev4-2}
\bibliography{references}

\begin{thebibliography}{20}
\bibitem [{Mat()}]{Math}%
  \BibitemOpen
  \href@noop {} {\bibinfo {title} {Mathematica{\texttrademark} 11.2.0.0
  computer package was used for the algebraic, numerical and graphics
  manipulations}}\BibitemShut {NoStop}%
\end{thebibliography}
\end{document}